\documentclass[preprint,showpacs,amsmath]{revtex4}
\usepackage{epsfig}

\def\ra{\rangle}
\def\la{\langle}

\def\Cb{{\Bbb C}}
\def\be{\begin{equation}}
\def\ee{\end{equation}}
\def\ba{\begin{array}}
\def\ea{\end{array}}
\def\Cb{\ \hbox{\vrule width 0.6pt height 6.5pt depth 0pt
              \hskip -3.2 pt} C}

\def\qed{\leavevmode\unskip\penalty9999 \hbox{}\nobreak\hfill
     \quad\hbox{\leavevmode  \hbox to.77778em{%
               \hfil\vrule   \vbox to.675em%
               {\hrule width.6em\vfil\hrule}\vrule\hfil}}
     \par\vskip3pt}
\newtheorem{theorem}{Theorem}

\newtheorem{lemma}[theorem]{Lemma}

\begin{document}

\title{Lower Bounds of Concurrence for Multipartite States}


\author{Xue-Na Zhu$^{1}$}
\author{Ming Li$^{2}$}
\author{Shao-Ming Fei$^{3}$}

\affiliation{$^1$Department of Mathematics, School of Science, South
China University of Technology, Guangzhou 510640, China\\
$^2$Department of Mathematics, School of Science, China
   University of Petroleum, 266555 Qingdao, China\\
$^3$School of Mathematical Sciences, Capital Normal
University, Beijing 100048, China}

\begin{abstract}
We study the entanglement of multipartite quantum
states. Some lower bounds of the multipartite
concurrence are reviewed. We further present more effective lower bounds
for detecting and qualifying entanglement, by
establishing functional relations between the concurrence and the generalized partial
transpositions of the multipartite systems.
\end{abstract}

\maketitle

\section{1.~ Introduction}
Entanglement is a distinctive feature of quantum mechanics, and an
indispensable ingredient in various kinds of quantum information
processing applications such as quantum computation \cite{di},
quantum teleportation \cite{teleportation}, dense coding
\cite{dense}, quantum cryptographic schemes \cite{schemes},
entanglement swapping \cite{swapping} and remote states preparation
(RSP) \cite{RSP1}. These effects based on quantum entanglement
have been demonstrated in many pioneering experiments.

An important theoretical challenge in the theory of quantum
entanglement is to give a proper description and quantification of
quantum entanglement for given quantum states. For bipartite quantum
systems, entanglement of formation (EOF) \cite{eof} and concurrence
\cite{concurrence,anote} are two well defined quantitative measures
of quantum entanglement. For two-qubit systems it has been proved
that EOF is a monotonically increasing function of the concurrence
and an elegant formula for the concurrence was derived analytically
by Wootters \cite{wotters}. However with the increasing dimensions
of the subsystems the computation of EOF and concurrence become
formidably difficult. A few explicit analytic formulae for EOF and
concurrence have been found only for some special symmetric states
\cite{Terhal-Voll2000,fjlw,fl,fwz,Rungta03}.

The first analytic lower bound of concurrence
that can be tightened by numerical optimization over some parameters
was derived in \cite{167902}.
In \cite{Chen-Albeverio-Fei1,chen} analytic
lower bounds on EOF and concurrence for any dimensional mixed
bipartite quantum states have been presented
by using the positive partial
transposition (PPT) and realignment separability criteria.
These bounds are exact for some
special classes of states and can be used to detect many bound
entangled states. In \cite{breuer} another lower bound on EOF for
bipartite states has been presented from a new separability
criterion \cite{breuerprl}. A lower bound of concurrence based on
local uncertainty relations (LURs) criterion is derived in
\cite{vicente}. This bound is further optimized in \cite{zhang}.
In \cite{edward,ou} the authors presented
lower bounds of concurrence for bipartite systems in terms of
a different approach. It has been shown that this lower bound has a
close relationship with the distillability of bipartite quantum states.
In Ref. \cite{X. S. Li} an explicit analytical lower bound of concurrence is
obtained by using positive maps, which is better than the ones in Refs. \cite{chen,breuer} in detecting
some quantum entanglement. These bounds give rise to a good quantitative estimation of
concurrence. They are supplementary in detecting quantum entanglement for bipartite systems.

When referring to multipartite systems, we focus on multipartite
concurrence, since the EOF is only defined for bipartite systems.
With the increasing of the number of quantum systems, quantifying
multipartite entanglement has become a much difficult task and only
few results are obtained. In this paper, we first give a brief
review of the lower bounds for multipartite concurrence in section
2. We present some new lower bounds of multipartite
concurrence in sections 3-5. These new bounds give rise to better
estimations of multipartite concurrence and are more effective in
detecting multipartite entanglement. Conclusions and remarks are
given in section 6.

\section{2.~ Lower bounds of multipartite concurrence}

We first recall the definition and some lower bounds of the multipartite
concurrence. Let ${\cal {H}}_{i}$, $i=1,...,N$, be Hilbert spaces with $d_i$ dimensions.
The concurrence of an $N$-partite state $|\psi\ra\in {\cal {H}}_{1}\otimes{\cal
{H}}_{2}\otimes\cdots\otimes{\cal {H}}_{N}$ is defined by
\cite{multicon}
\begin{eqnarray}\label{xxxx}
C_{N}(|\psi\ra\la\psi|)=2^{1-\frac{N}{2}}\sqrt{(2^{N}-2)-\sum_{\alpha}{\rm
Tr}[\rho_{\alpha}^{2}]},
\end{eqnarray}
where $\alpha$ labels all different reduced density matrices.

Up to constant factor (\ref{xxxx}) can be also expressed in another
way. Set $d_i=d, i=1,2,...,N$. The $N$-partite pure state $|\psi\ra$
is generally of the form,
\begin{eqnarray}\label{purestate}
|\psi\ra=\sum\limits_{i_{1},i_{2},\cdots,
i_{N}=1}^{d}a_{i_{1},i_{2},\cdots, i_{N}}|i_{1},i_{2},\cdots,
i_{N}\ra,\quad a_{i_{1},i_{2},\cdots, i_{N}}\in \Cb,
\end{eqnarray}
with $\sum\limits_{i_{1},i_{2},\cdots,i_{N}=1}^{d}
a_{i_{1},i_{2},\cdots, i_{N}}a_{i_{1},i_{2},\cdots, i_{N}}^\ast=1$.

Let $\alpha$ and $\alpha^{'}$ (resp.$\beta$ and $\beta^{'}$) be
subsets of the subindices of $a$, associated to the same sub Hilbert
spaces but with different summing indices. $\alpha$ (or
$\alpha^{'}$) and $\beta$ (or $\beta^{'}$) span the whole space of
the given sub-indix of $a$. The generalized concurrence of
$|\psi\ra$ is then given by \cite{anote},
\begin{eqnarray}\label{defmulticon}
C_{d}^{N}(|\psi\ra)=\sqrt{\frac{d}{2m(d-1)}\sum\limits_{p}
\sum\limits_{\{\alpha,\alpha^{'},\beta,\beta^{'}\}}^{d}
|a_{\alpha\beta}a_{\alpha^{'}\beta^{'}}-a_{\alpha\beta^{'}}a_{\alpha^{'}\beta}|^{2}},
\end{eqnarray}
where $m=2^{N-1}-1$, $\sum\limits_{p}$ stands for the summation over
all possible combinations of the indices of $\alpha$ and $\beta$.

For a mixed multipartite quantum state,
$\rho=\sum_{i}p_{i}|\psi_{i}\ra\la\psi_{i}|\in{\cal
{H}}_{1}\otimes{\cal {H}}_{2}\otimes\cdots\otimes{\cal {H}}_{N}$,
the corresponding concurrence is given by the convex roof:
\begin{eqnarray}\label{defe}
C_{N}(\rho)=\min_{\{p_{i},|\psi_{i}\}\ra}\sum_{i}p_{i}C_{N}(|\psi_{i}\ra).
\end{eqnarray}

In \cite{gao} the lower bound of concurrence for tripartite systems
has been studied by exploring the connection between the generalized
partial transposition criterion and concurrence. Let ${\cal{H}}_A$, ${\cal{H}}_B$
and ${\cal{H}}_C$ be three finite dimensional Hilbert spaces
associated with the subsystems $A$, $B$ and $C$, with dimensions
$\dim A=m$, $\dim B=n$ and $\dim C=p$. Define that $T_{r_k}$ (resp.
$T_{c_k}$), $k=A,B,C,AB,BC,AC$ to be the row (resp. column)
transpositions with respect to the subsystems $k$. Consider three
classes: 1) $y_i= \{c_k, r_k\}$, where $i=1,2,3$ for $k=A,B,C$
respectively; 2) $y_4= \{c_A, r_{BC}\}$, $y_5= \{c_{AB}, r_{C}\}$, $y_6=
\{c_{AC}, r_{B}\}$; 3) $y_7=\{c_A, r_B\}$, $y_8=\{c_A, r_C\}$, $y_9=\{c_B, r_C\}$.

For any $m\otimes n\otimes p$ $(m\leq n, p)$ tripartite mixed quantum
state $\rho$, the concurrence $C(\rho)$ defined in (\ref{xxxx})
satisfies
\begin{eqnarray*}\label{thgao}
&&C_{N}(\rho)\\
&&\geq\max\{\sqrt{\frac{1}{m(m-1)}}(||\rho^{T_{y_a}}||-1),
\sqrt{\frac{1}{n(n-1)}}(||\rho^{T_{y_b}}||-1),\sqrt{\frac{1}{r(r-1)}}(||\rho^{T_{y_c}}||-1)\}.
\end{eqnarray*}
where $q=\min(n,mp)$ and $r=\min(p,mn), y_a=y_1$ or $y_4, y_b=y_2$
or $y_6, y_c=y_3$ or $y_5$.

In \cite{mintert,aolita} the definition of multipartite concurrence
defined in (\ref{xxxx}) is re-expressed as
$C(|\psi\ra)=\sqrt{\la\psi|\otimes\la\psi|
A|\psi\ra\otimes|\psi\ra})$, with
$A=4(P_+-P_+^{(1)}\otimes\cdots\otimes P_+^{(N)})$. $P_+$ (resp. $P_-$) is the
projector ont o the globally symmetric (reps. antisymmetric) space. The
authors have obtained that the multipartite concurrence satisfies
\begin{eqnarray*}\label{thmintert}
[C_N(\rho)]^2\geq Tr(\rho\otimes\rho V),\end{eqnarray*} with
$V=4(P_+-P_+^{(1)}\otimes\cdots\otimes P_+^{(N)}-(1-2^{1-N})P_-)$.

In \cite{zhang,mintert,aolita}, it is shown that the multipartite
concurrence defined in (\ref{xxxx}) satisfies
\begin{eqnarray}\label{upperlowerboundo}
C_{N}(\rho)\geq \sqrt{(4-2^{3-N}){\rm
Tr}\{\rho^{2}\}-2^{2-N}\sum_{\alpha}{\rm Tr}\{\rho_{\alpha}^{2}\}}.
\end{eqnarray}

We derived an effective lower bound for
multipartite quantum systems in \cite{jpa}. First for tripartite case,
\begin{theorem}\label{th1} For an arbitrary $d\times d\times d$ mixed state $\rho$ in
${\cal{H}}\otimes {\cal{H}}\otimes {\cal{H}}$, the concurrence
$C(\rho)$ defined in (\ref{defmulticon}) satisfies
\begin{eqnarray}\label{11o}
\tau_{3}(\rho)\equiv\frac{d}{6(d-1)}\sum_{\alpha}^{\frac{d^{2}(d^{2}-1)}{2}}\sum_{\beta}^{\frac{d(d-1)}{2}}
[({C_{\alpha\beta}^{12|3}(\rho)})^{2}
+({C_{\alpha\beta}^{13|2}(\rho)})^{2}+({C_{\alpha\beta}^{23|1}(\rho)})^{2}]\leq C^{2}(\rho),
\end{eqnarray}
where $\tau_{3}(\rho)$ is a lower bound of $C(\rho)$,
\begin{eqnarray}
C_{\alpha\beta}^{12|3}(\rho)=\max\{0,\lambda(1)_{\alpha\beta}^{12|3}-\lambda(2)_{\alpha\beta}^{12|3}
-\lambda(3)_{\alpha\beta}^{12|3}-\lambda(4)_{\alpha\beta}^{12|3}\},
\end{eqnarray}
$\lambda(1)_{\alpha\beta}^{12|3}, \lambda(2)_{\alpha\beta}^{12|3},
\lambda(3)_{\alpha\beta}^{12|3}, \lambda(4)_{\alpha\beta}^{12|3}$
are the square roots of the four nonzero eigenvalues, in decreasing
order, of the non-Hermitian matrix
$\rho\widetilde{\rho}_{\alpha\beta}^{12|3}$ with
$\widetilde{\rho}_{\alpha\beta}^{12|3}=S_{\alpha\beta}^{12|3}\rho^{*}S_{\alpha\beta}^{12|3}$.
$C_{\alpha\beta}^{13|2}(\rho)$ and $C_{\alpha\beta}^{23|1}(\rho)$
are defined in a similar way to $C_{\alpha\beta}^{12|3}(\rho)$.
\end{theorem}

Theorem $\ref{th1}$ can be directly generalized to arbitrary multipartite
case.
\begin{theorem}\label{th2} For an arbitrary $N$-partite state $\rho\in
{\cal{H}}\otimes {\cal{H}}\otimes...\otimes{\cal{H}}$, the concurrence defined
in ($\ref{defmulticon}$) satisfies:
\begin{eqnarray}\label{tau}
\tau_{N}(\rho)\equiv\frac{d}{2m(d-1)}\sum_{p}\sum_{\alpha\beta}(C_{\alpha\beta}^{p}(\rho))^{2}\leq
C^{2}(\rho),
\end{eqnarray}
where $\tau_{N}(\rho)$ is the lower bound of $C(\rho)$,
$\sum\limits_{p}$ stands for the summation over all possible
combinations of the indices of $\alpha,\beta$,
$C_{\alpha\beta}^{p}(\rho)=\max\{0,
\lambda(1)_{\alpha\beta}^{p}-\lambda(2)_{\alpha\beta}^{p}
-\lambda(3)_{\alpha\beta}^{p}-\lambda(4)_{\alpha\beta}^{p}\}$,
$\lambda(i)_{\alpha\beta}^{p}$, $i=1, 2, 3, 4$, are the square roots
of the four nonzero eigenvalues, in decreasing order, of the
non-Hermitian matrix $\rho\widetilde{\rho}_{\alpha\beta}^{p}$ where
$\widetilde{\rho}_{\alpha\beta}^{p}=S_{\alpha\beta}^{p}\rho^{*}S_{\alpha\beta}^{p}$.
\end{theorem}

In \cite{rmp} we further obtained lower bound of multipartite concurrence by bipartite
partitions of the whole quantum systems.
For a pure N-partite quantum state $|\psi\ra\in {\mathcal
{H}}_{1}\otimes{\mathcal {H}}_{2}\otimes\cdots\otimes{\mathcal
{H}}_{N}$, $dim {\mathcal {H}}_{i}=d_i$, $i=1,...,N$, the concurrence of bipartite
decomposition between subsystems
$12\cdots M$ and $M+1\cdots N$ is defined by
\begin{eqnarray}\label{xx}
C_{2}(|\psi\ra\la\psi|)=\sqrt{2(1-{\rm Tr}\{\rho_{{1}{2}\cdots
{M}}^{2}\})},
\end{eqnarray}
where $\rho_{{1}{2}\cdots {M}}^{2}={\rm Tr}_{{M+1}\cdots
{N}}\{|\psi\ra\la\psi|\}$ is the reduced density matrix of
$\rho=|\psi\ra\la\psi|$ by tracing over the subsystems $M+1\cdots{N}$.

For a mixed multipartite quantum state,
$\rho=\sum_{i}p_{i}|\psi_{i}\ra\la\psi_{i}| \in {\mathcal
{H}}_{1}\otimes{\mathcal {H}}_{2}\otimes\cdots\otimes{\mathcal
{H}}_{N}$, the corresponding concurrence of (\ref{xx}) is then given
by the convex roof:
\begin{eqnarray}\label{def1}
C_{2}(\rho)=\min_{\{p_{i},|\psi_{i}\}\ra}\sum_{i}p_{i}C_{2}(|\psi_{i}\ra\la\psi_{i}|),
\end{eqnarray}
which will be called the bipartite concurrence.

The relation between the concurrences in (\ref{defe}) and the
bipartite concurrence in (\ref{def1}) can be directly given by
the following theorem.
\begin{theorem} For a multipartite quantum state $\rho\in
{\mathcal {H}}_{1}\otimes{\mathcal
{H}}_{2}\otimes\cdots\otimes{\mathcal {H}}_{N}$ with $N\geq 3$, the
following inequality holds,
\begin{eqnarray}\label{obound}
C_{N}(\rho)\geq\max 2^{\frac{3-N}{2}}C_{2}(\rho),
\end{eqnarray}
where the maximum is taken over all kinds of bipartite concurrence.
\end{theorem}

In terms of the lower bounds of bipartite concurrence derived from
PPT, realignment of the density matrix, local uncertainty relation
and the covariance matrix separability criterion in
\cite{chen,vicente,zhang}, and (\ref{obound}), we get the following theorem.
\begin{theorem} For any N-partite quantum state $\rho$, we have:
\begin{eqnarray}\label{newlowerbound}
C_{N}(\rho)\geq2^{\frac{3-N}{2}}\max\{B1,B2,B3\},
\end{eqnarray}
where
\begin{eqnarray*}
B1&=&\max_{\{i\}}\sqrt{\frac{2}{M_{i}(M_{i}-1)}}\left[\max(||{\mathcal
{T}}_{A}(\rho^{i})||,||R(\rho^{i})||)-1\right],\\
B2&=&\max_{\{i\}}\frac{2||C(\rho^{i})||- (1-{\rm
Tr}\{(\rho^{i}_{A})^{2}\})-(1-{\rm Tr}\{(\rho^{i}_{B})^{2}\})}
{\sqrt{2M_{i}(M_{i}-1)}},\\
B3&=&\max_{\{i\}}\sqrt{\frac{8}{M_{i}^{3}N_{i}^{2}(M_{i}-1)}}
(||T(\rho^{i})||-\frac{\sqrt{M_{i}N_{i}(M_{i}-1)(N_{i}-1)}}{2}),
\end{eqnarray*}
$\rho^i$ are all possible bipartite decompositions of $\rho$,
$M_{i}=\min{\{d_{s_{1}}d_{s_{2}}\cdots d_{s_{m}},
d_{s_{m+1}}d_{s_{m+2}}\cdots d_{s_{N}}\}}$,
$N_{i}=\max{\{d_{s_{1}}d_{s_{2}}\cdots d_{s_{m}},
d_{s_{m+1}}d_{s_{m+2}}\cdots d_{s_{N}}\}}$.
\end{theorem}

\section{3.~ Improved lower bounds of the multipartite concurrence}
In this section, we will derive a new bound for multipartite quantum
systems by using the following lemma.

\begin{lemma} For a bipartite density matrix $\rho\in H_A\otimes
H_B$. one has \cite{zhang}
\begin{equation}\label{5}
1-Tr\{\rho_{AB}^2\}\geq(1-Tr\{\rho_{A}^2\})-(1-Tr\{\rho_{B}^2\}),
\end{equation}
\begin{equation}\label{6}
1-Tr\{\rho_{AB}^2\}\geq(1-Tr\{\rho_{B}^2\})-(1-Tr\{\rho_{A}^2\}),
\end{equation}
where $\rho_{A|B}=Tr_{A}\{\rho_{B}\}$, $\rho_{B}=Tr_{B|A}\{\rho_{A}\}$.
\end{lemma}

\begin{theorem} For a multipartite quantum state $\rho\in
H_1\otimes H_2\otimes ... \otimes H_N$ with $N\geq3$, the following
inequality holds:
\begin{equation}\label{7}
C_N(\rho)\geq\max_{\{M=1,2,...,N-1\}}\left\{\left(2^{\frac{1-N}{2}}\sqrt{2^{N-M}+2^{M}-2}\right)C_2(\rho_M)\right\},
\end{equation}
where the maximum takes over all kinds of bipartite concurrences.
\end{theorem}

{\bf Proof.} For a pure multipartite state $ \vert\varphi\rangle\in
{\cal {H}}_{1}\otimes{\cal {H}}_{2}\otimes\cdots\otimes{\cal {H}}_{N}$,
one has $Tr\{\rho_{12...M}^2\}=Tr\{\rho_{M+1...N}^2\}$ for all $M={1,2,...,N-1}$.

From (\ref{5}) and (\ref{6}), we obtain
\begin{equation}
1-Tr{\rho_{12...Mi_1...i_p}^2}\geq(1-Tr{\rho_{12...M}^2})-(1-Tr{\rho_{i_1...i_p}^2}),
\end{equation}
and
\begin{equation}
1-Tr{\rho_{j_1...j_qM+1...N}^2}\geq(1-Tr{\rho_{M+1...N}^2})-(1-Tr{\rho_{j_1...j_q}^2}),
\end{equation}
where $M+1\leq i_1<...<i_p\leq N$, $p\leq N-M-1$ and $1\leq
j_1<...<j_q\leq M$, $q\leq M-1$.

From the above inequalities,  we have
\begin{eqnarray*}
C_N^2(\vert\varphi\rangle
\langle\varphi\vert)&&=2^{2-N}\left[(2^N-2)-\sum_\alpha
Tr{\rho_\alpha^2}\right]
=2^{2-N}\left(\sum_{k=1}^{2^N-2}(1-Tr{\rho_k^2)}\right)\\[1mm]
&&\geq2^{2-N}\left\{(2^{N-M}-1)(1-Tr{\rho_{12...M}^2})+(2^M-1)(1-Tr{\rho_{M+1...N}^2})\right\}\\[2mm]
&&=2^{2-N}\left\{(2^{N-M}+2^M-2)(1-Tr{\rho_{12...M}^2})\right\}\\[2mm]
&&=2^{2-N}\left\{(2^{N-M}+2^M-2)\frac{C_2(|\varphi\rangle_M\langle\varphi|)}{2}\right\},
\end{eqnarray*}
i.e. $C_N(\vert\varphi\rangle \langle\varphi\vert)\geq\max_{\{M=1,
2,..., N-1\}} \left(2^{\frac{1-N}{2}}\sqrt{2^{N-M}+2^{M}-2}\right)
C_2(\vert\varphi\rangle_M \langle\varphi\vert).$

Assuming that
 $\rho=\sum_{i} p_i\vert\varphi_i\rangle \langle\varphi_i\vert$ attains the minimal decomposition of the multipartite concurrence, one has
\begin{eqnarray*}
&&C_N(\rho)=\sum_{i}p_iC_N(\vert\varphi_i\rangle \langle\varphi_i\vert)\\
&&\qquad\geq2^{\frac{1-N}{2}}\sqrt{2^{N-M}+2^{M}-2}\sum_ip_iC_2(\vert\varphi_i\rangle_M\langle\varphi_i\vert)\\
&&\qquad\geq2^{\frac{1-N}{2}}\sqrt{2^{N-M}+2^{M}-2}\min_{\{p_i,\vert\varphi_i\rangle\}}\sum_ip_iC_2(\vert\varphi_i\rangle_M\langle\varphi_i\vert)\\
&&\qquad=\left(2^{\frac{1-N}{2}}\sqrt{2^{N-M}+2^{M}-2}\right)C_2(\rho_M).
\end{eqnarray*}
Therefore we have
$$
C_N(\rho)\geq\max_{\{M=1,2,...,N-1\}}\left\{\left(2^{\frac{1-N}{2}}\sqrt{2^{N-M}+2^{M}-2}\right)C_2(\rho_M)\right\}.
$$\qed

\section{4.~ Functional relations between concurrence and the generalized partial transpositions}

Let us consider an $N$-qubit state, the generalized $W$ state,
\begin{equation}\label{11}
\vert\varphi\rangle=a_1\vert10...0\rangle+a_2\vert01...0\rangle+...+a_N\vert00...1\rangle.
\end{equation}

\begin{theorem} For any $N$-qubit mixed state with
decomposition on the generalized $W$ states, $\rho=\sum_{i}
p_i\vert\varphi_i\rangle \langle\varphi_i\vert$, such that
$\vert\varphi_i\rangle$ can be written in the form (\ref{11}) for
all $i$, the concurrence $C(\rho)$ satisfies

\begin{equation}\label{12}
C(\rho)\geq2^{1-\frac{N}{2}}\max
\left\{|\rho^{T_{\Gamma^1_\alpha}}|\
-1,\max_{M}\left\{\sqrt{\frac{2^{N-M}+2^{M}-2}{4}}(|
\mathcal{R}_{\Gamma^1_\alpha\vert\Gamma^2_\alpha}(\rho)|\
-1)\right\}\right\},
\end{equation}
where $\Gamma^1_\alpha$, $\Gamma^2_\alpha$ denote two subsets of the
indices $\{1, 2, ..., N\}$,
$\Gamma^1_\alpha\cap\Gamma^2_\alpha=\emptyset$,
$\Gamma^1_\alpha\cup\Gamma^2_\alpha=\{1, 2, ..., N\}, \alpha=1, ...,
d$, $M=(1,2,...,N-1)$  is the number of elements of $\Gamma^1_\alpha$.
\end{theorem}

{\bf Proof.} \ An $N$-qubit $W$ state can be viewed as $d$ different
bipartite systems. From the results for bipartite systems
\cite{chen}, these $d$ bipartite separations give rise to,
respectively
$$
1-Tr\{\rho^2_{\Gamma^1_\alpha}\}\geq\frac{1}{2}(|
\mathcal{R}_{\Gamma^1_\alpha\vert\Gamma^2_\alpha}(\rho)|\ -1)^2,
\alpha=1, ..., d.
$$

Hence
\begin{eqnarray*}
C(\vert\varphi\rangle \langle\varphi\vert)&&=2^{1-\frac{N}{2}}
\sqrt{d-\sum_{\alpha=1}^{d} Tr\{\rho^2_{\Gamma^1_{\alpha}}\}}\\
&&=2^{1-\frac{N}{2}}\sqrt{\frac{2d-\sum_{\alpha=1}^d Tr\{\rho^2_{\Gamma^1_\alpha}\}-\sum_{\alpha=1}^d Tr\{\rho^2_{\Gamma^2_\alpha}\}}{2}}\\
&&\geq2^{1-\frac{N}{2}}\max_{M}\sqrt{\frac{2^{N-M}+2^{M}-2}{2}\left(1-Tr\{\rho^2_{\Gamma^1_\alpha}\}\right)}\\
&&\geq2^{1-\frac{N}{2}}\max_{M}\sqrt{\frac{2^{N-M}+2^{M}-2}{4}}(|
\mathcal{R}_{\Gamma^1_\alpha\vert\Gamma^2_\alpha}(\rho)|\ -1).
\end{eqnarray*}

Let $\rho=\sum_{i} p_i\vert\varphi_i\rangle \langle\varphi_i\vert$
attain the minimal decomposition of the multipartite concurrence. Note
that $| \mathcal{R}(\rho)|\leq\sum_{i} p_i|
\mathcal{R}(\vert\varphi_i\rangle
\langle\varphi_i\vert)|$ \cite{chen}. One has
\begin{eqnarray*}
&&C(\rho)=\sum_{i}p_iC(\vert\varphi_i\rangle \langle\varphi_i\vert)\\
&&\geq 2^{1-\frac{N}{2}}\max_{M}\sqrt{\frac{2^{N-M}+2^{M}-2}{4}}
\sum_{i}p_i(| \mathcal{R}_{\Gamma^1_\alpha\vert\Gamma^2_\alpha}(\vert\varphi_i\rangle \langle\varphi_i\vert)|\ -1)\\
&&\geq 2^{1-\frac{N}{2}}\max_{M}\sqrt{\frac{2^{N-M}+2^{M}-2}{4}}(|
\mathcal{R}_{\Gamma^1_\alpha\vert\Gamma^2_\alpha}(\rho)|\ -1),
\end{eqnarray*}
From which one gets (\ref{12}).
\qed

\section{5.~ Entanglement detecting and estimation of concurrence}

In this section, we use the above several
lower bounds of multipartite concurrence to detect quantum entanglement. We will show by
examples that these bounds provide a better estimation of the
multipartite concurrence.

{\sf{(1) Lower bound and separability}}

An N-partite quantum state $\rho$ is fully separable if and only if
there exist $p_{i}$ with $p_{i}\geq0, \sum\limits_{i}p_{i}=1$ and
pure states $\rho_{i}^{j}=|\psi_{i}^{j}\ra\la\psi_{i}^{j}|$ such
that
$\rho=\sum_{i}p_{i}\rho_{i}^{1}\otimes\rho_{i}^{2}\otimes\cdots\otimes\rho_{i}^{N}.$
It is easily verified that for a fully separable multipartite state
$\rho$, $\tau_{N}(\rho)$ defined in (\ref{tau}) is zero. Thus
$\tau_{N}(\rho)>0$ indicates that there must be some kinds of
entanglement inside the quantum state, which shows that the lower
bound $\tau_{N}(\rho)$ can be used to recognize entanglement.

As an example we consider a tripartite quantum state \cite{acin},
$\rho=\frac{1-p}{8}I_{8}+p|W\ra\la W|$, where $I_{8}$ is the
$8\times8$ identity matrix, and
$|W\ra=\frac{1}{\sqrt{3}}(|100\ra+|010\ra+|001\ra)$ is the
tripartite W-state. By using the generalized correlation matrix
criterion presented in \cite{hassan} the entanglement of $\rho$ is
detected for $0.3068 < p \leq 1$. From our theorem, we have that the
lower bound $\tau_{3}(\rho)>0$ for $0.2727 < p \leq 1$. Therefore
our bound detects entanglement better in
this case. If we replace W with GHZ state in $\rho$, the criterion
in \cite{hassan} detects the entanglement of $\rho$ for $0.35355 < p
\leq 1$, while $\tau_{3}(\rho)$ detects, again better, the
entanglement for $0.2 < p \leq 1$.

{\sf{(2) Estimation of multipartite concurrence}}

The lower bounds together with some upper bounds can be used to
estimate the value of the concurrence. In
\cite{zhang,mintert,aolita}, it is shown that the upper and lower
bound of multipartite concurrence satisfy
\begin{eqnarray}\label{upperlowerbound}
\sqrt{(4-2^{3-N}){\rm Tr}\{\rho^{2}\}-2^{2-N}\sum_{\alpha}{\rm
Tr}\{\rho_{\alpha}^{2}\}}\leq
C_{N}(\rho)\leq\sqrt{2^{2-N}[(2^{N}-2)-\sum_{\alpha}{\rm
Tr}\{\rho_{\alpha}^{2}\}]}.
\end{eqnarray}

In fact we can obtain a more effective upper bound for multi-partite
concurrence. Let $\rho=\sum\limits_{i}\lambda_{i}|\psi_{i}\ra\la
\psi_{i}|\in {\mathcal
{H}}_{1}\otimes{\mathcal{H}}_{2}\otimes\cdots\otimes{\mathcal
{H}}_{N}$, where $|\psi_{i}\ra$s are the  orthogonal pure states and
$\sum\limits_{i}\lambda_{i}=1$. We have
\begin{eqnarray}\label{newupperbound}
C_{N}(\rho)=\min_{\{p_{i},|\varphi_{i}\}\ra}\sum_{i}p_{i}C_{N}(|\varphi_{i}\ra\la\varphi_{i}|)
\leq\sum_{i}\lambda_{i}C_{N}(|\psi_{i}\ra\la\psi_{i}|).
\end{eqnarray}

We now show that our upper and lower bounds can be better than that
in $(\ref{upperlowerboundo})$ by detailed examples.

\begin{figure}[tbp]
\resizebox{10cm}{!}{\includegraphics{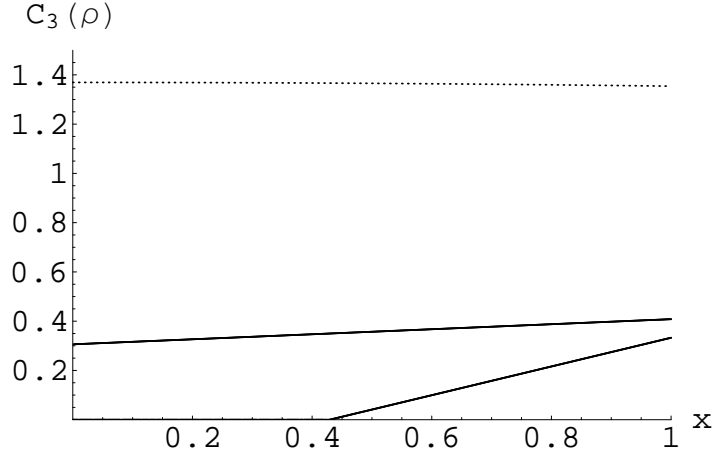}}
\caption{Our
lower and upper bounds of $C_{3}(\rho)$ from
(\ref{newlowerbound}) and (\ref{newupperbound}) (solid line) and the
upper bound obtained in (\ref{upperlowerbound}) (dot line) while
the lower bound in (\ref{upperlowerbound}) is always zero.
\label{fig1}}
\end{figure}

{\it{Example 1:}} Consider the $2\times 2\times 2$ D\"ur-Cirac-Tarrach states defined by \cite{dur}:
$$\rho=\sum_{\sigma=\pm}\lambda_{0}^{\sigma}|\psi_{0}^{\sigma}\ra\la
\psi_{0}^{\sigma}|+\sum_{j=1}^{3}\lambda_{j}(|\psi_{j}^{+}\ra\la\psi_{j}^{+}|+|\psi_{j}^{-}\ra\la\psi_{j}^{-}|),
$$
where the orthonormal Greenberger-Horne-Zeilinger (GHZ)-basis
$|\psi_{j}^{\pm}\ra\equiv\frac{1}{\sqrt{2}}(|j\ra_{12}|0\ra_{3}\pm|(3-j)\ra_{12}|1\ra_{3})$,
$|j\ra_{12}\equiv|j_{1}\ra_{1}|j_{2}\ra_{2}$ with $j=j_{1}j_{2}$ in
binary notation. From theorem 2 we have that the lower bound of
$\rho$ is $\frac{1}{3}$. If we mix the state with white noise,
$\rho(x)=\frac{(1-x)}{8}I_{8}+x\rho,$ by direct computation we have,
as shown in FIG. $\ref{fig1}$, the lower bound obtained in
$(\ref{upperlowerboundo})$ is always zero, while the lower bound in
$(\ref{newlowerbound})$ is larger than zero for $0.425\leq x\leq 1$,
which shows that $\rho(x)$ is detected to be entangled at this
situation. And the upper bound (dot line) in
$(\ref{upperlowerboundo})$ is much larger than the upper bound we
have obtained in $(\ref{newupperbound})$ (solid line).

Actually, our new lower bound in (\ref{7}) is different from the
lower bound in (\ref{obound}), which can be seen from the following
example.

{\it{Example 2:}} Consider the generalized GHZ state:
$\vert\varphi\rangle=\cos\theta\vert00...0\rangle+\sin\theta\vert11...1\rangle.$
It is easy to obtain that
$Tr\rho^2_{i_1,i_2,...,i_m}=1-2\sin^2{\theta}\cos^2{\theta}$ for all
$i_1\not=i_2\not=...\not=i_m\in{1,2,...,N}$. Hence we have by
definition
$C(\vert\varphi\rangle)=2^{1-\frac{N}{2}}\sqrt{(2^N-2)(2\sin^2{\theta}\cos^2{\theta})}.$
By our new lower bound in (\ref{7}), we get
\begin{eqnarray*}
C_N(\rho)&&\geq\max_{\{M=1,2,...,N-1\}}\left\{\left(2^{\frac{1-N}{2}}\sqrt{2^{N-M}+2^{M}-2}\right)C_2(\rho_M)\right\}\\
&&=\max_{\{M=1,2,...,N-1\}}\left\{\left(2^{\frac{1-N}{2}}\sqrt{2^{N-M}+2^{M}-2}\right)\sqrt{4\sin^2{\theta}\cos^2{\theta}}\right\}\\
\end{eqnarray*}

For example, $N=4$, we get $C_N(\vert\varphi\rangle
\langle\varphi\vert)=\sqrt{7 \sin^2{\theta}\cos^2{\theta}}$. From our
bound we have $C_N(\vert\varphi\rangle
\langle\varphi\vert)\geq\sqrt{4\sin^2{\theta}\cos^2{\theta}}>\sqrt{2\sin^2{\theta}\cos^2{\theta}}$,
where $\sqrt{2\sin^2{\theta}\cos^2{\theta}}$ is the bound from \cite{rmp}.

\section{6.~ Remarks and conclusions}

By establishing functional relations between the concurrence and the generalized partial
transpositions of the multipartite systems, we have presented some effective lower bounds
for detecting and qualifying entanglement for multipartite systems.
These bounds can be also served as
separability criteria. They detect entanglement of some states
better than some separability criteria.

Generally, to derive a lower
bound of multipartite concurrence, we calculate the multipartite
concurrence for pure states first. Then by using the convex property
of the quantities in the calculation one can directly find a tight
lower bound. Mintert et al. in \cite{mintert7} have derived a
precise lower bound for bipartite concurrence, which detects mixed
entangled states with a positive partial transpose. It would be
interesting and challenging to use this approach to derive a lower
bound for multipartite concurrence.

\bigskip
\noindent{\bf Acknowledgments}\, This work is supported by the NSFC 10875081 and PHR201007107.

\end{document}